\documentclass[%
a4paper,
superscriptaddress,
showpacs,
 amsmath,amssymb,
 aps,
10pt,twocolumn,pre,
]{revtex4-1}

 \usepackage{xcolor}
\usepackage{graphics}

\colorlet{mylinkcolor}{blue!66!black!80}
\usepackage[colorlinks=true,linkcolor=mylinkcolor,citecolor=mylinkcolor,filecolor=cyan,urlcolor=mylinkcolor,breaklinks=true]{hyperref}

\usepackage{dsfont}

\usepackage[utf8]{inputenc}

\newcommand{\avg}[1]{\langle#1\rangle}

\newcommand{\dd}{{\rm d}}
\newcommand{\e}{{\rm e}}

\newcommand{\Det}{\text{det}}
\newcommand{\ic}{{\rm i}}
\newcommand{\doubleavg}[1]{\langle\!\langle#1\rangle\!\rangle}
\renewcommand{\L}{\mathsf{L}}
\newcommand{\B}{{\rm B}}

\begin{document}
\title{Optimal performance of periodically driven, stochastic heat engines\\
under limited control}

\author{Michael Bauer}
\affiliation{%
II. Institut für Theoretische Physik, Universität Stuttgart, 70550 Stuttgart, Germany
}%
\author{Kay Brandner}%
\affiliation{%
Department of Applied Physics, Aalto University, FI-00076 Aalto, Finland
}%
 \author{Udo Seifert}%
\affiliation{%
II. Institut für Theoretische Physik, Universität Stuttgart, 70550 Stuttgart, Germany
}%
\date{\today}

\begin{abstract}
We consider the performance of periodically driven 
stochastic heat engines in the linear response regime. 
Reaching the theoretical bounds for efficiency and efficiency
at maximum power typically requires full control over the design
and the driving of the system. We develop a framework which allows
to quantify the role that limited control over the system has on the
performance. Specifically, we show that optimizing the driving
entering the work extraction for a given temperature protocol leads
to a universal, one-parameter dependence for both maximum efficiency and maximum 
power as a function of efficiency.
In particular, we show that reaching
Carnot efficiency (and, hence, Curzon-Ahlborn efficiency at maximum power)
requires to have control over the amplitude of
the full Hamiltonian of the system.
Since the kinetic energy cannot be controlled by an external parameter,
heat engines based on underdamped 
dynamics can typically not reach Carnot efficiency. We illustrate our general theory 
with a paradigmatic case study of a heat engine consisting of an 
underdamped charged particle in a modulated 
two-dimensional harmonic trap in the presence of a magnetic field. 
\end{abstract}
\maketitle

\section{Introduction}

As a consequence of the first and second law of thermodynamics,
the efficiency of any device converting heat into work is subject to the Carnot bound
\begin{align}
 \eta_\text{C}\equiv 1-{T_c}/{T_h},
\end{align}
where $T_c$ and $T_h$ are the temperatures of the cold and hot bath, respectively.
This constraint holds both within macroscopic \cite{call85} and stochastic thermodynamics \cite{seif12}.
In either of these realms, it can be saturated under ideal reversible conditions,
i.e., if all crucial variables of the system can be controlled and the driving is 
performed adiabatically. 
Under realistic conditions, however, the Carnot bound is typically unattainable. 
Sources of irreversibility are, for instance, heat leaks, friction in the working 
medium, and constraints on the equation of state in macroscopic 
\cite{Gordon1992,Chen1994} and microscopic engines
\cite{Hondou2000,JimenezdeCisneros2008,Zhang2010,Izumida2015b}.
The efficiency of thermoelectric devices is typically reduced due to heat transfer
by phonons and imperfect electronic transmission functions \cite{Mahan1996,Heremans2008}.

Besides efficiency, power constitutes a second crucial measure for the performance of a 
heat engine.
Maximizing both of these figures simultaneously is generally expected to be impossible, since power inevitably vanishes in the adiabatic limit, where Carnot efficiency can be realized. 
However,  Benenti {\sl et al} \cite{Benenti2011} pointed out that this dilemma might, in principle, be overcome in systems with broken time-reversal symmetry. 
Furthermore, Allahverdyan {\sl et al} argued that high efficiency at finite power is indeed feasible in a generalized finite-time Carnot cycle under certain conditions
 \cite{Allahverdyan2013}.
On the other hand, within large classes of thermoelectric and Brownian heat engines, 
additional constraints have recently been discovered, which rule out the option of 
Carnot efficiency at finite power, regardless of how the system behaves under time-revesal \cite{Brandner2013,Brandner2013a,Brandner2015b,Brandner2015a}.

A common way to avoid these intricacies is to consider efficiency at maximum power as a benchmark parameter. 
Within the endoreversible setup, which takes into account irreversible heat exchange between
the reservoirs and an otherwise ideal Carnot engine, Curzon and Ahlborn showed that this 
figure is given by \cite{curz75,Salamon2001}
\begin{align}\label{CA}
 \eta_{\text{CA}}\equiv 1-\sqrt{{T_c}/{T_h}}. 
\end{align}
Remarkably, like the Carnot bound, $\eta_{{{\rm CA}}}$ depends only on the temperatures of the reservoirs but not the properties of the engine itself. 
Nevertheless, it has meanwhile turned out that efficiency at maximum power does not admit a universal bound but rather depends crucially on the admissible space of control parameters \cite{VandenBroeck2005,Schmiedl2008a,Tu2008,JimenezdeCisneros2007,Esposito2009,Izumida2009,Esposito2010,Sanchez-Salas2010,seif11a,izum15}.
A systematic and quantitative description of how limited control affects the performance of a
heat engine is, however, currently not available. 

In stochastic thermodynamics, Brownian particles are ideal model systems to investigate such fundamental aspects.
Generally, heat engines based on Brownian dynamics can be divided into two classes.
First, systems featuring directed current of particles in spatially periodic temperature profiles \cite{Hondou2000,Asfaw2004,Berger2009},
and second, Brownian particles in a periodically modulated trapping potential \cite{Schmiedl2008a,Izumida2010,abre11,baue12,Tu2014,Dechant2015,Dechant2016}.
Specifically, a minimal heat engine consisting of a single particle in a harmonic trap, which is alternately coupled to two heat baths of different
temperature, has been introduced theoretically in Ref. \cite{Schmiedl2008a}.
This set-up has later been realized experimentally in two different variants \cite{blic12,mart15,mart16} and miniaturized even further down to the scale of single 
ions \cite{abah12,ross15}. 

For thermoelectric engines in the linear response regime and without a magnetic field, the conditions to saturate the bounds on efficiency are well understood. 
Specifically, the electric and the heat current have to be tightly coupled \cite{Kedem1965,VandenBroeck2005}, i.e., proportional to each other.
This requirement can be fulfilled if the transmission of electrons is restricted 
to a narrow energy-band \cite{Humphrey2002,Humphrey2005}. 
A $\delta$-shaped transmission function thus constitutes the thermoelectric analog of 
quasistatic driving protocols, which allow cyclic engines to operate reversibly. 
Indeed, it has been shown that the work and the heat fluxes in a periodically driven Brownian system satisfy the tight-coupling criterion in the adiabatic limit \cite{Brandner2015a}.
An analogous result holds for discrete systems with periodically modulated energy levels \cite{Proesmans2015a,Proesmans2015c}.

Breaking the time-reversal symmetry of thermoelectric engines and thus seemingly improving 
their performance \cite{Benenti2011,Stark2014} requires the presence of an external magnetic field. 
In periodic engines, this symmetry can be broken more easily by choosing driving 
protocols, which are not invariant under time-reversal \cite{Brandner2015a}.
Including an additional magnetic field in such systems can still be beneficial.
For example, high efficiency at maximum power for energy transfer in magnetic field coupled oscillator networks is found in Ref. \cite{Sabass2015}.
However, in order to describe the Lorentz-force in a thermodynamically consistent way,
the momenta of the system must be fully taken into account \cite{Hondou2000,Jimenez-aquino2008,Gomez-Marin2008,Rana2014,Tu2014}. 
These kinetic degrees of freedom can typically not be controlled directly but rather provide only an additional source of dissipation. 
Underdamped dynamics thus intrinsically allows only ''limited control''.


In this article, using the linear response formalism developed in Ref. \cite{Brandner2015a}, we explore the finite-time performance of periodically driven, stochastic heat engines
under limited control.
Within a general framework, we maximize power under well defined limitations.
For systems without magnetic field and for a large class of systems with magnetic field,
we find that detailed balance implies a certain symmetry of the correlation functions,
which simplifies the thermodynamic analysis of the engine.
In fact, we show that, under these conditions, the benchmark parameters of a generic cyclic heat 
engine can be expressed in terms of a single figure of merit. 
This result is in strong analogy to the standard theory used for thermoelectric devices \cite{bell08,Goupil2011}.

The paper is organized as follows.
In section \ref{sec:OC}, we 
present the linear response framework and explain how the power output of heat engines is maximized for fixed efficiency.
Section \ref{sec:limitedcontrol_optdriving} is devoted to the concept of limited control and the calculation of maximal power output at fixed efficiency. 
In section \ref{sec:eigenfunctions}, this limitation is related to the figure of merit
by assuming that the control function can be expressed as a finite sum of eigenfunctions of the adjoint Fokker-Planck operator.
We illustrate our findings with a case study
in section \ref{sec:example_schmiedl}
and conclude in section \ref{sec:conclusion}.

\section{Periodically driven, stochastic heat engines}
\label{sec:OC}

\begin{figure}
\includegraphics{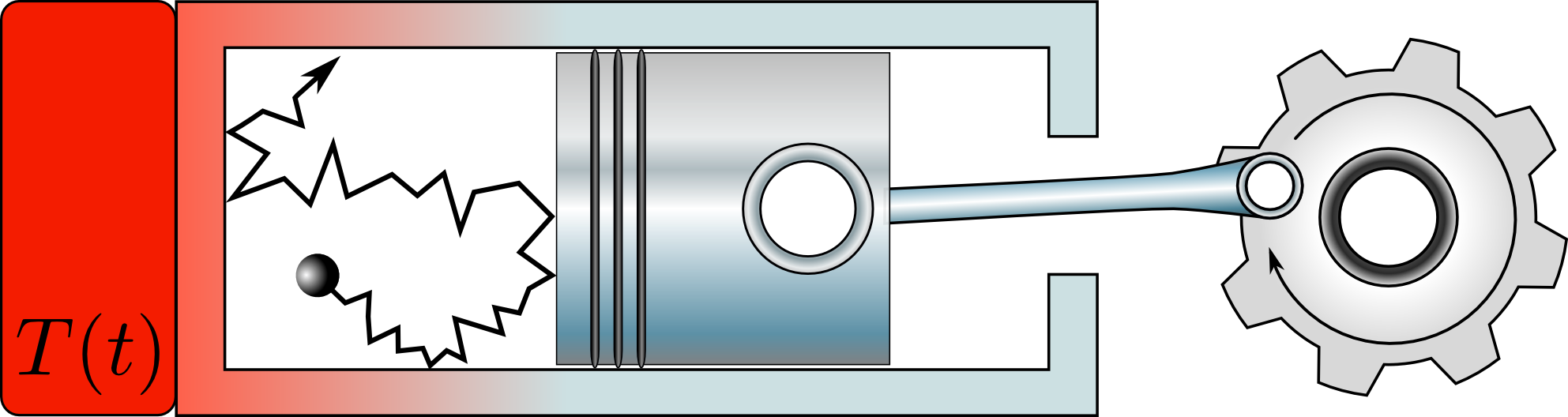}
 \caption{Scheme of a cyclic Brownian heat engine.
A stochastic working medium depicted as a colloidal particle is
coupled to a thermal reservoir with time-dependent temperature $T(t)$.
Useful work can be extracted from the medium by varying certain
external control parameters, e.g., the position of a moving piston,
according to properly chosen protocols.}
\label{fig:heatengine}
\end{figure}

\subsection{Set-up}

The heat engines investigated here consist of several parts: the working medium, the heat bath,
and a component for work exchange (see figure \ref{fig:heatengine}). The working medium is of stochastic nature, which is, for instance, a single particle following Langevin dynamics.
The temperature of the heat bath, which surrounds the working medium, changes periodically,
thus one heat bath is playing the role of both the hot and cold bath. 
The energy of the working medium is modulated periodically by external control parameters, 
allowing to extract work.
Specifically, we follow the linear response theory for periodically driven, stochastic heat engines
developed in Ref. \cite{Brandner2015a}.
The system with phase space variables ${\bf x}$ undergoes a modulation of energy $H({\bf x},t)$ and temperature $T(t)$,
which are 
both $\mathcal{T}$-periodic in time $t$.
The temperature
\begin{align}
 T(t)&=T+\Delta T \gamma_q(t),
\label{eq:Temperature}
\end{align}
is varied by the temperature protocol $\gamma_q(t)$,
with $0\leq \gamma_q(t) \leq1$, and small amplitude $\Delta T>0$. The lower and upper bound correspond to what is usually the cold and hot bath.
The Hamiltonian
\begin{align}
H({\bf x},t)&=H_0({\bf x})+ \Delta H   g_w({\bf x},t),
\label{eq:energy_modulation}
\end{align}
is varied by
the work protocol $g_w({\bf x},t)$ and small amplitude $\Delta H$ around its equilibrium value $H_0({\bf x})$.
For given temperature protocol $\gamma_q(t)$ and fixed system parameters,
a heat engine is realized by a suitable choice of the work protocol $g_w({\bf x},t)$,
such that work is extracted in a cycle from the heat bath.
Throughout the paper, subscript $w$ and $q$ refer to work and heat, respectively.
The dimensionless work protocol $ g_w({\bf x},t)$ and the temperature protocol $\gamma_q(t)$ are crucial for the performance of the engine,
since they contain the information how the energetics changes in time.
In the context of stochastic thermodynamics, work and heat are defined as \cite{seif12}
\begin{align}
 \dot W(t)&\equiv -\int \dd^d {\bf x} \dot H ({\bf x,t}) p({\bf x,t}),\\
 \dot Q(t)&\equiv \int \dd^d {\bf x} H ({\bf x,t}) \dot  p({\bf x,t}),
\end{align}
where $p({\bf x},t)$ is the phase space distribution of the working medium.
In the periodic state, the entropy production reads
\begin{align}
 \dot S=-\frac{1}{\mathcal{T}}\int_0^\mathcal{T} \dd t \frac{\dot Q(t)}{T(t)} \geq 0,
\end{align}
which is positive due to the second law.
In irreversible thermodynamics the entropy production is expressed as \cite{Brandner2015a}
\begin{align}
 \dot S=J_w \mathcal{F}_w+J_q \mathcal{F}_q,
\end{align}
where the fluxes are identified as
\begin{align}
 J_w&=\frac{1}{\mathcal{T}}\int_0^\mathcal{T} \int \dd^d {\bf x}\dot g_w({\bf x},t)  p({\bf x,t}),\\
 J_q&=\frac{1}{\mathcal{T}}\int_0^\mathcal{T} \int \dd^d {\bf x} \gamma_q(t) H({\bf x},t) \dot p({\bf x,t}),
\end{align}
and affinities 
\begin{align}
 \mathcal{F}_w=\Delta H/T,&    &\mathcal{F}_q=\Delta T/T^2.
\label{eq:Affinities}
\end{align}

\subsection{Fokker-Planck dynamics}
The dynamics of the phase space distribution of the working medium is governed by a Fokker-Planck equation
\begin{align}
 \dot p({\bf x},t)=\L({\bf x},t)p({\bf x},t),
\end{align}
where the dot denotes the derivative with respect to time.
After a transient time, the system reaches a periodic steady state $p^c({\bf x},t)=p^c({\bf x},t+\mathcal{T})$,
to which we restrict our analysis.
In equilibrium, the distribution and the dynamics are described by the zero-order terms
\begin{align}
 p^c({\bf x},t)|_{\Delta T=\Delta H=0}&=p^\text{eq}({\bf x})=\exp[-H_0({\bf x})/ T]/Z_0,\\
 \L({\bf x},t)|_{\Delta T=\Delta H=0}&=\L_0({\bf x}),
\end{align}
with normalization $Z_0$ and Boltzmann's constant set to $1$ throughout.

For systems obeying Fokker-Planck dynamics, the detailed balance condition reads
\begin{align}
 \L_0 ({\bf x}) p_{\text{eq}}({\bf x})=p_{\text{eq}}({\bf x})\L_0^\dagger ({\boldsymbol \varepsilon \bf x}),
\label{eq:detailed_balance}
 \end{align}
which constitutes an operator equation.
It connects the Fokker-Planck operator and its adjoint operator $\L_0^\dagger$
and includes the behavior of variables which change their sign under time reversal \cite{Risken1989}.
Here the transformation
$\boldsymbol \varepsilon$ leaves positions untouched, inverts momenta, and
includes the inversion of an external magnetic field.

\subsection{Explicit form of the Onsager coefficients}
The Onsager coefficients for work and heat
connect fluxes
\begin{align}
 J_w&=L_{ww}\mathcal{F}_w+L_{wq}\mathcal{F}_q,\\
 J_q&=L_{qw}\mathcal{F}_w+L_{qq}\mathcal{F}_q,
\end{align}
and affinities in the linear response regime.
Then the entropy production reads
\begin{align}
 \dot S &=\sum_{\alpha,\beta=w,q}L_{\alpha \beta}\mathcal{F}_\alpha \mathcal{F}_\beta.
\label{eq:entropyproduction}
\end{align}
Since the form above is quadratic, positivity requires
\begin{align}
&L_{qq}, L_{ww}\geq 0,& 
 \frac{(L_{wq}+L_{qw})^2}{4L_{ww}L_{qq}} \leq 1&.
\label{eq:secondlaw_OC}
\end{align}
Equality in the latter condition leads to the
tight-coupling regime.

For systems modulated in the way described above, the Onsager coefficients have recently been derived in Ref. \cite{Brandner2015a}.
With the equilibrium average
\begin{align}
 \langle A \rangle_{\bf x}
 &\equiv   \int \dd^d {\bf x} \ A({\bf x},t) \ p^\text{eq}({\bf x}),
\end{align}
and
the combined average over a period of operation and phase space
\begin{align}
 \doubleavg{A}\equiv \frac{1}{\mathcal{T}}\int_0^\mathcal{T} \dd t \langle A \rangle_{\bf x},
\end{align}
the Onsager coefficients are found to be functionals of the $g_\alpha ({\bf x},t)$ 
\begin{align}
 L_{\alpha\beta}&=L^{\text{ad}}_{\alpha\beta}+L^{\text{dyn}}_{\alpha\beta},\nonumber\\
 L^{\text{ad}}_{\alpha\beta}&\equiv-\doubleavg{\delta \dot g_\alpha({\bf x},t) \delta g_\beta({\bf x},t)},\nonumber\\
L^{\text{dyn}}_{\alpha\beta}&\equiv \int_0^\infty \dd \tau \doubleavg{\delta \dot g_\beta({\bf x},t-\tau)   \exp(\L_0^\dagger \tau)  \delta \dot g_\alpha({\bf x},t)},
\label{eq:general_OC}
\end{align}
with $\alpha,\beta=w,q$,
and the deviation from the average
\begin{align}
\delta A\equiv A -\avg{A}_{\bf x}.
\label{eq:def_delta}
\end{align}
Inside the brackets we omit the phase space variables for simplicity.
For a uniform notation we have defined
\begin{align}
 g_q({\bf x},t)&\equiv-\gamma_q(t) H_0({\bf x})
 \label{eq:gq}
.\end{align}

\subsection{Maximal power for fixed efficiency}
Optimizing the performance of the engine for given temperature protocol $\gamma_q(t)$ and fixed system parameters ($H_0({\bf x}),T,\Delta T,\Delta H,\mathcal{T},\mathrm{L}_0^\dagger$), we search the work protocol $g_w=g_w({\bf x},t)$ which maximizes the power output functional
\begin{align}
P[g_w]\equiv -T J_w \mathcal{F}_w,
\end{align}
at fixed efficiency.
By using the rescaled power
\begin{align}
\mathcal{P}[g_w]\equiv{P[g_w]}/{T\mathcal{F}_q^2}
= - (\chi^2 L_{ww}+\chi L_{wq}), 
\label{eq:P}
\end{align}
and rescaled heat flux
\begin{align}
\mathcal{J}_q[g_w]\equiv{J_q[g_w]}/{{\mathcal{F}}_q}= \chi L_{qw}+  L_{qq},
\label{eq:Jq}
\end{align}
with $\chi\equiv\mathcal{F}_w/\mathcal{F}_q$,
the maximized power output at fixed efficiency does not depend on $\Delta T$ and $\Delta H$.
In principle, changing the work protocol changes both the power output and the efficiency
\begin{align}
\eta[g_w]\equiv{{P[g_w]}}/{{J}_q[g_w]}.
\label{eq:def_eff_functional}
\end{align}
However, for a systematic investigation of the performance it is advantageous to keep efficiency fixed.
As a constraint,
we thus demand the rescaled efficiency
\begin{align}
\bar\eta\equiv{\mathcal{P}[g_w]}/{\mathcal{J}_q[g_w]}={\eta[g_w]}/{\eta_\text{C}}\leq 1,
\label{eq:def_eff}
\end{align}
to be constant, where we have used
Carnot efficiency $\eta_\text{C}=\Delta T/ T=T \mathcal{F}_q$ and (\ref{eq:Affinities},\ref{eq:P},\ref{eq:Jq}).
We thus have to maximize the
objective functional
\begin{align}
 \mathcal{P}&[g_w,\Lambda]\equiv \mathcal{P}[g_w] +\Lambda (\bar \eta \mathcal{J}_q[g_w] - \mathcal{P}[g_w]) \nonumber\\
=&(\Lambda-1)\chi^2 L_{ww}+(\Lambda-1)\chi L_{wq}+\Lambda \bar\eta \chi L_{qw}+\Lambda \bar\eta L_{qq},
\label{eq:objective_functional}
\end{align}
with respect to the time dependent phase space function $g_w({\bf x},t)$.
This objective functional contains the power output (\ref{eq:P}) and the constraint for fixed efficiency (\ref{eq:def_eff}),
where $\Lambda$ is the Lagrange multiplier.

\begin{figure}
\includegraphics{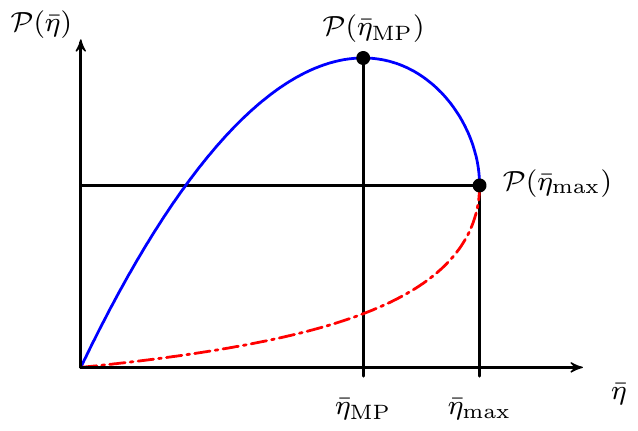}
\caption{Schematic performance curve $\mathcal{P}(\bar\eta)$ of a heat engine.
Symbols are explained in the main text.}
\label{fig:performance_curve}
\end{figure}

The maximization of (\ref{eq:objective_functional}) at fixed $\bar\eta$ and $\Lambda$ yields the optimal work protocol $g_w^*(\Lambda)$,
and, by insertion into (\ref{eq:def_eff}), we obtain the corresponding Lagrange multiplier $\Lambda(\bar\eta)$.
Thus, at given efficiency, we calculate the maximal power 
\begin{align}
\mathcal{P}(\bar\eta)\equiv \mathcal{P}[ g_w^*,\Lambda(\bar\eta)],
\label{eq:performance_curve}
\end{align}
with $g_w^*=g_w^*(\Lambda(\bar\eta))$.
Then, for given system parameters and temperature protocol
we are able to investigate the performance of a heat engine under optimal driving as a function of efficiency.
We calculate maximal efficiency $\bar\eta_\text{max}$ and efficiency at maximum power $\bar\eta_\text{MP}$
with corresponding power $\mathcal{P}(\bar\eta_\text{max})$ and $\mathcal{P}(\bar\eta_\text{MP})$, respectively.
The power output is maximal if the constraint (\ref{eq:def_eff}) is ignored,
i.e., by setting $\Lambda=0$, from which the definition
\begin{align}
\mathcal{P}(\bar\eta_\text{MP})\equiv \mathcal{P}[ g_w^*(0),0]
\label{eq:def_maxPower}
\end{align}
follows. These characteristic points are shown on a schematic performance curve in figure \ref{fig:performance_curve}.
For realistic, macroscopic heat engines
typically loop-shaped performance curves $\mathcal{P}(\bar\eta)$ are obtained \cite{Gordon1992,Chen1994}.
Likewise, they appear for a Brownian heat engine with consideration of kinetic energy \cite{Zhang2010}
and for certain heat engines that do not fulfill the tight-coupling condition \cite{JimenezdeCisneros2008,Izumida2015b}.
We focus on the upper branch of the loop since it contains maximum power and maximum efficiency.
The lower branch corresponds to a non-optimal protocol, leading to lower power output.

Furthermore, we are interested whether or not the bound on power found in Ref. \cite{Brandner2015b,Brandner2015a}
\begin{align}
 \mathcal{P}(\bar\eta)\leq 4 \bar\eta (1-\bar\eta) \mathcal{P}_0 \leq \mathcal{P}_0,
\label{eq:bound_on_power}
\end{align}
can be saturated.
We call
\begin{align}
\mathcal{P}_0&\equiv-\frac{1}{4}\doubleavg{\delta g_q L_0^\dagger \delta g_q} 
\label{eq:normalization_constant}
,\end{align}
the global power bound.

\section{Thermodynamics of limited control under optimal driving}
\label{sec:limitedcontrol_optdriving}

\subsection{Definition of limited control and Fourier representation of the Onsager coefficients}
\label{sec:limited_control}

Having set the preliminaries for the maximization of the power output at given efficiency in the previous section,
we now define limited control in our setup. 
The work protocol defined in (\ref{eq:energy_modulation}) is assumed to consist of $n$ contributing terms
\begin{align}
g_w({\bf x},t)&=\sum_{i=1}^{n} \gamma_{w_i}(t) g_{w_i}({\bf x}),
 \label{eq:gw_zerlegung}
\end{align}
where the control functions $g_{w_i}({\bf x})$ are fixed and the $\gamma_{w_i}(t)$ are the control parameters, which are varied externally.
Then, the Hamiltonian can be divided into two contributions
\begin{align}
 H_0({\bf x})&=H_0^c({\bf x})+H_0^{r}({\bf x}), \label{eq:H0_control}\\
H_0^c({\bf x})&=\sum_{i=1}^n (H_0)_i g_{w_i}({\bf x}).\nonumber
\end{align}
The first term is spanned by the control functions, where $(H_0)_i$ are the respective coefficients.
The remaining second term is not influenced by the control functions.
Only if $H_0^r({\bf x})=0$, we speak of "full control".
If, for example, the strength $k(t)$ of a harmonic trap can be modulated,
we have $n=1$ and $g_{w}({\bf x})=  k(t) x^2/2$, where $x$ is the position of the particle.
If this system is underdamped, we also consider the kinetic energy $H_0^r({\bf x})=mv^2/2$,
which is not controlled by the trap.
Here $m$ is the particle mass and $v$ its velocity.

Since the control parameters and the temperature protocol
are $\mathcal{T}$-periodic, they
can be written in terms of their Fourier components 
\begin{align}
 \gamma_{w_i,q}(t) &=\sum_{k=-\infty}^\infty c_k^{w_i,q} \e^{\ic   k\Omega  t},\\
c_k^{w_i,q}&\equiv \frac{1}{\mathcal T} \int_0^{\mathcal T}   \gamma_{w_i,q}(t) \e^{-\ic   k\Omega t }\dd t,
\label{eq:cn}
\end{align}
with $\Omega\equiv2\pi/\mathcal{T}$.
Using (\ref{eq:gw_zerlegung}) the Onsager coefficients (\ref{eq:general_OC}) take the form
\begin{align}
 L_{qq}&=\sum_{k=1}^\infty k^2 \Omega^2 c_{-k}^q \hat{C}[\delta H_0,\delta H_0,k]c_k^q ,\label{eq:OC_Fourier}\\
 L_{ww}&=\sum_{i,j=1}^{n}\sum_{k=1}^\infty k^2 \Omega^2 { c}_{-k}^{w_i}  \hat{C}[\delta g_{w_i},\delta g_{w_j},k] {c}_{k}^{w_j},\nonumber\\
 L_{wq}&=-\sum_{i=1}^{n}\sum_{k=-\infty}^\infty k^2 \Omega^2 { c}_{-k}^{w_i}  \tilde{C}[\delta g_{w_i},\delta H_0,k] {c}_{k}^{q},\nonumber\\
 L_{qw}&=-\sum_{i=1}^{n}\sum_{k=-\infty}^\infty k^2 \Omega^2 { c}_{-k}^{q}  \tilde{C}[\delta H_0,\delta g_{w_i},k] {c}_{k}^{w_i},\nonumber
\end{align}
where we have defined
\begin{align}
 \hat{C}[R,S,k]&\equiv \int_{-\infty}^{\infty} \e^{-\ic k \Omega t} C[R,S,t] \dd t,\\
\tilde{C}[R,S,k]&\equiv \ic \langle R S\rangle_{\bf x}/k \Omega+ \int_{0}^{\infty} \e^{-\ic k \Omega t} C[R,S,t] \dd t,
\end{align}
with the correlation function \cite{Risken1989}
\begin{align}
C[R,S,t>0]&\equiv \langle S \e^{\L_0^\dagger t} R \rangle_{\bf x},\label{eq:def_corr}\\
C[R,S,t<0]&\equiv C[S,R,-t].\nonumber
\end{align}
From these definitions, two useful 
properties follow
\begin{align}
\hat{C}[R,S,k]&= \hat{C}[S,R,-k],\label{eq:Ctilde_symm}\\
\hat{C}[R,S,k]&=\tilde{C}[R,S,k]+\tilde{C}[S,R,-k],\nonumber
\end{align}
where we need to change the variable of integration $t\to - t$ to show the latter equality.

\subsection{Optimal protocol for maximum power for given efficiency}
\label{sec:general_opt_protocol}

We now maximize the objective functional (\ref{eq:objective_functional}) 
with respect to the Fourier coefficients $c^{w_j}_k$. We use the Onsager coefficients in the form (\ref{eq:OC_Fourier}),
and split sums with $k=\pm 1,\pm 2,\dots$ into two parts with $k=1,2,\dots$, which takes into account the complex nature of the $c^{w_j}_k$.
This procedure yields the optimal protocol $g^*_w(\Lambda)$ in terms of the Fourier coefficients of the $\gamma_{w_j}(t)$
\begin{multline}
 (c_k^{w_j})^*(\Lambda)=\frac{c_k^q}{\chi} \sum_{i=1}^{n} \hat{C}^{-1}_{ji}(k) \Big[ \tilde{C}[\delta g_{w_i},\delta H_0,k]\\-\frac{\bar\eta\Lambda}{1-\Lambda}\tilde{C}[\delta H_0,\delta g_{w_i},-k]\Big],
\label{eq:general_opt_prot}
\end{multline}
where the Lagrange multiplier $\Lambda$ is still to be determined.
Here,
$\hat{C}^{-1}_{ji}(k)$ is the inverse matrix of $\hat{C}[\delta g_{w_i},\delta g_{w_j},k]$
with $\hat{C}^{-1}_{ij}(k)=\hat{C}^{-1}_{ji}(-k)$.
Using the optimal protocol, the heat current (\ref{eq:Jq}) and the objective functional (\ref{eq:objective_functional}) 
become
\begin{multline}
\;\;\;\;\;\;\,\mathcal{J}_q[g^*_w(\Lambda)]=2 \mathcal{P}(\bar\eta_{\text{MP}})+\mathcal{J}_q^\text{idle}+\frac{2\bar\eta\Lambda}{1-\Lambda}\mathcal{D},\label{eq:Jq_P_withLambda}\\
\mathcal{P}[g^*_w(\Lambda),\Lambda]=(1-\Lambda+2\Lambda\bar\eta)\mathcal{P}(\bar\eta_{\text{MP}})+\Lambda\bar\eta \mathcal{J}_q^\text{idle}\\
+\frac{\bar\eta^2\Lambda^2}{1-\Lambda}\mathcal{D}.
\end{multline}
The crucial point is that
the quantities on the left hand side depend on
three terms $\mathcal{P}(\bar\eta_{\text{MP}})$, $\mathcal{J}_q^\text{idle}$, and $\mathcal{D}$, which we introduce now.

First, $\mathcal{P}(\bar\eta_{\text{MP}})$ is the maximum power defined in (\ref{eq:def_maxPower}),
which is found to be
\begin{multline}
\mathcal{P}(\bar\eta_\text{MP})=
\sum_{k=1}^\infty k^2 \Omega^2 |c_k^q|^2 \times\\
\sum_{i,j=1}^{n} 
\tilde{C}[\delta g_{w_i},\delta H_0,-k] \hat{C}^{-1}_{ij}(k) \tilde{C}[\delta g_{w_j},\delta H_0,k],
\label{eq:res_Pmax}
\end{multline}
with $|c_k^q|^2=c_k^q c_{-k}^q$.
Alternatively, the maximum power can be written as
$\mathcal{P}(\bar\eta_{\text{MP}})=\chi^2 L_{ww}^*|_{\Lambda=0}\geq 0$,
which is positive due to the second law (\ref{eq:secondlaw_OC}), where $ L_{ww}^*$ is the Onsager coefficient evaluated with the optimal protocol.

Second, the
corresponding heat current at maximum power can be written from (\ref{eq:Jq}) as
\begin{align}
 \mathcal{J}_q[g^*_w(0)]=2 \mathcal{P}(\bar\eta_{\text{MP}})+\mathcal{J}_q^\text{idle},
\label{eq:Jq_maxP}
\end{align}
with
\begin{multline}
\mathcal{J}_q^\text{idle}\equiv L_{qq}-
\sum_{k=-\infty}^\infty k^2 \Omega^2 |c_k^q|^2 \times\\
\sum_{i,j=1}^{n} 
\hat{C}[\delta g_{w_i},\delta H_0,-k] \hat{C}^{-1}_{ij}(k) \tilde{C}[\delta g_{w_j},\delta H_0,k].
\label{eq:def_Jqidle}
\end{multline}
A first hint on the physical meaning of $\mathcal{J}_q^\text{idle}$ follows by
using (\ref{eq:def_eff}), (\ref{eq:res_Pmax}) and (\ref{eq:Jq_maxP}).
We then find the efficiency at maximum power
\begin{align}
 \bar\eta_{\text{MP}}=\frac{\mathcal{P}(\bar\eta_\text{MP})}{2\mathcal{P}(\bar\eta_\text{MP}) +\mathcal{J}_q^\text{idle}}.
 \label{eq:EMP_ohneZT}
\end{align}
Obviously, this result coincides with the Curzon-Ahlborn value $\eta_{\text{CA}}=\eta_{\text{C}}/2$ only for $\mathcal{J}_q^\text{idle}=0$. 

Third, in the general case with a constraint on efficiency (\ref{eq:def_eff}),
i.e.,
$\Lambda\neq 0$, we additionally need
\begin{multline}
 \mathcal{D}\equiv 
\sum_{k=1}^\infty k^2 \Omega^2 |c_k^q|^2 \times\\
\sum_{i,j=1}^{n} 
\tilde{C}[\delta H_0,\delta g_{w_i},k] \hat{C}^{-1}_{ij}(k) \tilde{C}[\delta H_0,\delta g_{w_j},-k],
\end{multline}
to describe the maximal power and heat current (\ref{eq:Jq_P_withLambda}).
This term is similar to the maximum power (\ref{eq:res_Pmax}).

The maximal power output for periodically driven heat engines (\ref{eq:Jq_P_withLambda}) is thus expressed by these three terms.
The Lagrange multiplier has to be determined by considering the 
constraint (\ref{eq:def_eff}).
In the following section, we show that the characteristics of the efficiency-power curves can be derived 
without explicitly making use of the rather involved expressions above.

\subsection{Performance in terms of one figure of merit}
\label{sec:FoM}

\subsubsection{Single parameter description}
In this subsection, it is assumed that all correlation functions are symmetric
\begin{align}
 C[R,S,t]=C[S,R,t],
\label{eq:symm_C}
\end{align}
with $R({\bf x}),S({\bf x})=\delta H_0({\bf x}),\delta g_{w_i}({\bf x})$,
leading to $\mathcal{D}=\mathcal{P}(\bar\eta_\text{MP})$ in (\ref{eq:Jq_P_withLambda}).
In Appendix \ref{sec:detailed_balance}, we derive that this symmetry indeed follows from detailed balance (\ref{eq:detailed_balance}) for over- and underdamped systems without magnetic field
and for a large class of systems with magnetic field.
We now show that this symmetry allows to express both, maximum efficiency and efficiency at maximum power, in terms of the dimensionless figure of merit
\begin{align}
 {\mathcal{ZT}} \equiv \frac{4\mathcal{P}(\bar\eta_\text{MP})}{\mathcal{J}_q^\text{idle}}.
 \label{eq:def_ZT}
\end{align}
With this definition, (\ref{eq:EMP_ohneZT}) can be rewritten as
\begin{align}
 \bar\eta_{\text{MP}}=\frac{\mathcal{ZT}}{2\mathcal{ZT}+4}.
\label{eq:EMP}
\end{align}
Maximum power for arbitrary but fixed $\bar\eta$
follows by taking into account the constraint for fixed efficiency (\ref{eq:def_eff}) with (\ref{eq:Jq_P_withLambda}),
which yields a quadratic equation for the Lagrange multiplier $\Lambda$ 
\begin{align}
0&= \bar\eta \mathcal{J}_q[g_w^*(\Lambda)]-\mathcal{P}[g_w^*(\Lambda)] \\
 &={\mathcal{P}(\bar\eta_\text{MP})} \left[\frac{\bar\eta^2 }{(\Lambda-1)^2}-(1-\bar\eta)^2\right]+\bar\eta\mathcal{J}_q^\text{idle} 
\label{eq:LambdaEQ}
.\end{align}
Its solution
\begin{align}
 \Lambda_\pm(\bar\eta)=1\mp\frac{\bar\eta}{\sqrt{(1-\bar\eta)^2-4\frac{\bar\eta}{\mathcal{ZT}}}},
\label{eq:LagrangeMultiplier}
\end{align}
is real only
if efficiency is below the maximal efficiency
\begin{align}
 \bar\eta_\text{max}
=\frac{\sqrt{\mathcal{ZT}+1}-1}{\sqrt{\mathcal{ZT}+1}+1},
\label{eq:max_eff}
\end{align}
at which the Lagrange multiplier diverges, $|\Lambda_\pm(\bar\eta_\text{max})|\to \infty$.
Furthermore, we are interested in the power at maximum efficiency
\begin{align}
 \mathcal{P}(\bar\eta_\text{max})\equiv\mathcal{P}[g_w^*(\Lambda_\pm(\bar\eta_\text{max})),\Lambda_\pm(\bar\eta_\text{max})].
\label{eq:def_Petamax}
\end{align}
Using (\ref{eq:Jq_P_withLambda}), (\ref{eq:LagrangeMultiplier}) and (\ref{eq:max_eff}), 
we obtain our first main result 
\begin{align}
  \frac{\mathcal{P}(\bar\eta_\text{max})}{\mathcal{P}(\bar\eta_\text{MP})}
= 1-\bar\eta_\text{max}^2.
\label{eq:PowerAtMaxEff}
\end{align}
Finite power at maximum efficiency is thus only possible if the maximum efficiency is below the Carnot value $\bar\eta=1$.

Combining (\ref{eq:EMP}) and (\ref{eq:max_eff})
leads to a relation between the efficiency at maximum power and maximum efficiency
\begin{align}
 \bar\eta_\text{MP}=\frac{\bar\eta_\text{max}}{1+\bar\eta_\text{max}^2},
\label{eq:master_curve}
\end{align}
which constitutes our second main result. From Ref. \cite{Benenti2011} it is known that for steady state devices with broken time reversal symmetry
two parameters are needed to express maximum efficiency and efficiency at maximum power. 
In contrast, here these quantities are described by only one parameter, eventually leading to (\ref{eq:master_curve}).
We recall that in the framework from section \ref{sec:OC}, time reversal symmetry is easily broken by the protocols,
whereas in Ref. \cite{Benenti2011} there is no protocol and instead time reversal symmetry is broken by a magnetic field.

\begin{figure}
\includegraphics{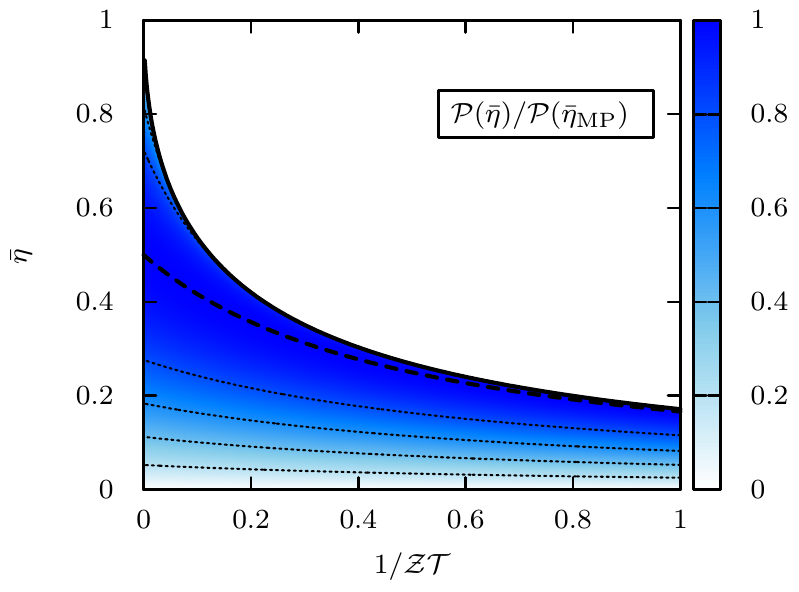}
\caption{Maximal power of periodically driven, stochastic heat engines ${\mathcal{P}(\bar\eta)}$ (\ref{eq:general_power})
normalized by its maximum ${\mathcal{P}(\bar\eta_\text{MP})}$ as a function of efficiency $\bar\eta$ and inverse figure of merit $1/{\mathcal{ZT}}$.
The solid black line at the top is the maximal efficiency $ \bar\eta_\text{max}$ (\ref{eq:max_eff}), above which the engine cannot operate.
Efficiency at maximum power $\bar\eta_{\text{MP}}$ (\ref{eq:EMP}) is marked with a dashed line. 
When the figure of merit becomes small, maximum efficiency and efficiency at maximum power both vanish.
When the figure of merit becomes large, maximal efficiency is Carnot efficiency and efficiency at maximum power is half the Carnot efficiency.
Along the dotted lines power is constant.}
\label{fig:general_power}
\end{figure}

\begin{figure}
\includegraphics{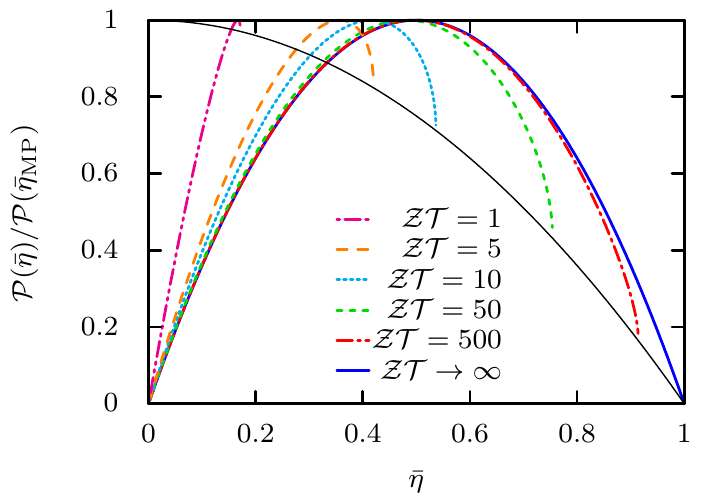}
\caption{Maximal power ${\mathcal{P}(\bar\eta)}/{\mathcal{P}(\bar\eta_\text{MP})}$ (\ref{eq:general_power}) as a function of efficiency $\bar\eta$,
for different values of the figure of merit ${\mathcal{ZT}}$. This figure shows a set of cuts through figure \ref{fig:general_power} for fixed $\mathcal{ZT}$. The power at maximal efficiency (\ref{eq:PowerAtMaxEff}) is given by the solid black line,
which vanishes at Carnot efficiency.}
\label{fig:general_power2}
\end{figure}

Now we show that the figure of merit $\mathcal{ZT}$ is also suitable to
describe the maximal power output for given efficiency.
The two solutions of the Lagrange multiplier (\ref{eq:LagrangeMultiplier})
result in two branches of the performance curve (\ref{eq:performance_curve})
\begin{align}
 \mathcal{P}_\pm(\bar\eta)\equiv\mathcal{P}[g_w^*(\Lambda_\pm(\bar\eta)),\Lambda_\pm(\bar\eta)],
\end{align}
from which
we choose the upper branch $\mathcal{P}(\bar\eta)=\mathcal{P}_+(\bar\eta)$,
since
\begin{align}
 \frac{\mathcal{P}_+(\bar\eta)-\mathcal{P}_-(\bar\eta)}{\mathcal{P}(\bar\eta_\text{MP})}=
4\bar\eta  \sqrt{(1-\bar\eta)^2-4\frac{\bar\eta}{{\mathcal{ZT}}}}\geq 0
\label{eq:branches_comparison}
.\end{align}
Collecting the results of this subsection, we rewrite (\ref{eq:Jq_P_withLambda}) for the maximal power at fixed efficiency of periodically driven, stochastic heat engines as
\begin{align}
 \frac{\mathcal{P}(\bar\eta)}{\mathcal{P}(\bar\eta_\text{MP})}
&=2\bar \eta \left(1-\bar\eta +\sqrt{(1-\bar\eta)^2-4\frac{\bar\eta}{\mathcal{ZT}}}+\frac{2}{\mathcal{ZT}} \right),
\label{eq:general_power}
\end{align}
where the right hand side depends only on the efficiency $\bar\eta$ and the figure of merit $\mathcal{ZT}$.
This universal expression
constitutes our third main result.
It is shown
color-coded in figure \ref{fig:general_power} as a function of the (inverse) figure of merit,
together with maximum efficiency (\ref{eq:max_eff}) and efficiency at maximum power (\ref{eq:EMP}).
Alternatively, the maximal power of the engine for given efficiency is shown in figure \ref{fig:general_power2} for different values of $\mathcal{ZT}$.
Loop-shaped performance curves similar to those of macroscopic real heat engines \cite{Gordon1992,Chen1994} are thus obtained.
The power at maximum efficiency (\ref{eq:PowerAtMaxEff}) is also shown.

The
result (\ref{eq:general_power})
contains all the information of the optimal performance of the engine at maximum power for given efficiency.
From the definition (\ref{eq:def_ZT}) with (\ref{eq:res_Pmax}) and (\ref{eq:def_Jqidle})
it can be seen that
the figure of merit is fixed by the temperature protocol and the system parameters.
Then $\mathcal{P}(\bar\eta)$ gives the output delivered by the engine
for a fixed efficiency $\bar\eta$
if the system is driven by the optimal protocol $g_w^*(\Lambda(\bar\eta))$.

\subsubsection{Carnot efficiency and the tight coupling regime}
\label{sec:tight_coupling}

An interesting limiting case is $\mathcal{ZT}\to \infty$,
or equivalently $\mathcal{J}_q^\text{idle}=0$, where maximal efficiency (\ref{eq:max_eff}) is the Carnot efficiency
and efficiency at maximum power (\ref{eq:EMP}) is the Curzon-Ahlborn value.
This limit corresponds to the endoreversible case 
in the context of macroscopic heat engines \cite{Gordon1992},
and to the overdamped case of Brownian heat engines \cite{Brandner2015a}.
Then the Lagrange multiplier (\ref{eq:LagrangeMultiplier}) reads
\begin{align}
\lim_{\mathcal{ZT}\to \infty} \Lambda_+(\bar\eta)=\frac{1-2\bar\eta}{1-\bar\eta},
\end{align}
leading to maximum power output
\begin{align}
\lim_{\mathcal{ZT}\to\infty}
 \frac{\mathcal{P}(\bar\eta)}{\mathcal{P}(\bar\eta_\text{MP})}
=4\bar\eta (1-\bar\eta)
\label{eq:power_full_control}
,\end{align}
from (\ref{eq:general_power}).
In this limiting case the efficiency-power diagram is no longer loop-shaped, but a parabola (see figure \ref{fig:general_power2}).

If all functions contained in $H_0({\bf x})$ can be controlled by the work protocol (\ref{eq:gw_zerlegung}),
i.e., $H_0^r({\bf x})=0$ in (\ref{eq:H0_control}),
it follows by using (\ref{eq:Ctilde_symm})
that the idle heat flux $\mathcal{J}_q^\text{idle}$ (\ref{eq:def_Jqidle}) 
vanishes.
Then, for $\bar\eta=1$ and using (\ref{eq:Ctilde_symm}), the optimal protocol (\ref{eq:general_opt_prot})
reads 
\begin{align}
 (c_k^{w_j})^*(\Lambda)&=c_k^q/\chi \sum_{i=1}^{n} \hat{C}^{-1}_{ji}(k)  \hat{C}[\delta g_{w_i},\delta H_0,k],\\
&=c_k^q  (H_0)_j/\chi,
\end{align}
which is equivalent to
\begin{align}
   g_w^*({\bf x},t)={\gamma_q(t) } H_0({\bf x})/{\chi}.
\label{eq:Carnot_condition}
\end{align}
Therefore the engine operates at Carnot efficiency if the full amplitude of the Hamiltonian (\ref{eq:energy_modulation}) is modulated according to (\ref{eq:Carnot_condition}).

If control is limited, i.e., $H_0^r({\bf x})\neq 0$,
the expression for the idle heat flux $\mathcal{J}_q^\text{idle}$ (\ref{eq:def_Jqidle})
is rather involved,
but typically non-vanishing.
Therefore, we conclude that if the full Hamiltonian cannot be controlled,
Carnot efficiency (and Curzon-Ahlborn efficiency at maximum power)
can not be achieved,
which is, for instance, the case in underdamped systems.

These statements holds for all periodically driven, stochastic heat engines introduced in section \ref{sec:OC},
and constitute our fourth main result.
We note that if (\ref{eq:Carnot_condition}) is valid, the tight-coupling condition follows,
since directly from (\ref{eq:general_OC})
we find
$L_{wq}=L_{qw}=-\chi L_{ww}=-L_{qq}/\chi$, which leads to equality in the second expression in (\ref{eq:secondlaw_OC}).
In this case the Onsager matrix is found to be symmetric (see "reciprocity relations" in Ref. \cite{Brandner2015a}).

\subsubsection{Remarks on the figure of merit}

In this subsection we mention several important points regarding the figure of merit.
First, in our analysis, we have imposed $\mathcal{ZT}\geq 0$, or equivalently $\mathcal{J}_q^\text{idle}\geq 0$.
A negative figure of merit yields unphysical results, e.g., power larger than in the case of full control (\ref{eq:power_full_control}).
Second, there is a freedom of choice in the definition of the figure of merit,
which we have fixed in
(\ref{eq:def_ZT}) such that
maximum efficiency (\ref{eq:max_eff}) and efficiency at maximum power (\ref{eq:EMP}) take the same form
as for 
thermoelectric devices \cite{Benenti2011,Brandner2013}.
Additionally, adapting the results from Ref. \cite{Gordon1991} to our notation, we find that the thermoelectric figure of merit
\begin{align}
 ZT=4P(\eta_\text{MP})/\mathcal{F}_q \kappa,
\end{align}
has a similar form as (\ref{eq:def_ZT}).
The thermal conductivity $\kappa$ is responsible
for an inevitable heat current $\mathcal{F}_q \kappa$ through the thermoelectric device similar to $\mathcal{J}_q^\text{idle}$.
Third, we note that for given efficiency at maximum power, it is always possible to 
define a figure of merit through (\ref{eq:EMP}).
However, it is then not guaranteed 
that this choice is also suitable to express the maximum efficiency in the form (\ref{eq:max_eff}).
Indeed we have found a structure in analogy to thermoelectric devices,
where
typically the power is maximized with respect to one thermodynamic affinity. In contrast, we maximize the power output with respect to the full work protocol.
These facts hint that the concept of a figure of merit of the form (\ref{eq:def_ZT}) might be universal for optimal processes in the linear response regime.

\section{Limited control over eigenfunctions of the adjoint Fokker-Planck operator}
\label{sec:eigenfunctions}

\subsection{Expansion of the Onsager coefficients}
In the previous section we have found bounds on efficiency smaller than the Carnot value
in terms of the figure of merit.
To gain a deeper insight
into the connection between limited control and the figure of merit,
we assume that in the control function (\ref{eq:gw_zerlegung}) and the Hamiltonian (\ref{eq:H0_control}) the functions $g_{w_i}({\bf x})=\varphi_i^\dagger({\bf x})$ are eigenfunctions of the adjoint Fokker-Planck operator $\L_0^\dagger$, 
which simplifies the evaluation of the Onsager coefficients (\ref{eq:general_OC}).
The phase space variables are allowed to be even or odd with respect to the transformation $\boldsymbol \varepsilon$.
For simplicity, in this section,
we assume that no external magnetic field is present.
Note that the eigenfunctions are usually not symmetric in $\boldsymbol\varepsilon$,
in contrast to the control functions $g_{w_i}({\bf x})$.
How this apparent paradox is resolved is shown in Appendix \ref{sec:detailed_balance}.
Useful properties of the eigenfunctions of the adjoint Fokker-Planck operator
are shown in Appendix \ref{sec:app_eigenfunction}.

We assume that
in total $N$ eigenfunction are needed,
from which we can control $n$,
whereas $N-n$ are used to span the part of the Hamiltonian which cannot be controlled
\begin{align}
 H_0^r({\bf x})=\sum_{i=n+1}^N (H_0)_i \varphi_i^\dagger({\bf x}).
\end{align}
The components of $H_0 ({\bf x})$ in the basis of eigenfunctions are given by
\begin{align}
 (H_0)_i=  \int \dd^d {\bf x}   \varphi_i({\bf x})  H_0 ({\bf x}),
\end{align}
where $\varphi_i({\bf x})$ are the corresponding eigenfunctions of the (regular) Fokker-Planck operator.
They have
eigenvalues $-\lambda_i$ with $i=0,1,2,\dots$
For example, $\varphi_0({\bf x})=p_{\text{eq}}({\bf x})$ is the equilibrium distribution with eigenvalue $\lambda_0=0$.

A suitable example for the present scheme is a three-dimensional, anisotropic, overdamped Brownian heat engine,
which is an extension of the example given in Ref. \cite{Brandner2015a}.
From the results found therein,
it can easily be deduced that the required eigenfunctions for $i=1,2,3$ have the form $x^2-T/m\omega_x^2$, $y^2-T/m\omega_y^2$, and $z^2-T/m\omega_z^2$,
where $x,y,z$ are directions in space and $\omega_x,\omega_y,\omega_z$ are the frequencies of the harmonic trap in the respective direction.
These three eigenfunctions are needed to express $ H_0({\bf x})$,
since all coordinates appear in the Hamiltonian and, hence, $N=3$.
If, for instance, $\omega_z$ cannot be controlled, but, $\omega_{x,y}$ can,
the $z$-coordinate is not contained in $g_w({\bf x},t)$ and thus $n=2$.

With the eigenfunction expansion and the abbreviation
\begin{align}
 c_k^{q_i}\equiv - c_k^q (H_0)_i,
\end{align}
the Onsager coefficients (\ref{eq:general_OC})
take the form
\begin{align}
 L_{\alpha \beta}
=\sum_{k=-\infty}^\infty\sum_{i=1}^{N} \frac{-\ic k \Omega \lambda_i }{\lambda_i -\ic k \Omega} c_k^{\alpha_i} c_{-k}^{\beta_i} ,
\label{eq:OC_eigenfunctions}
\end{align}
for $\alpha,\beta=w,q$ and $c_k^{w_i}=0$ for $n<i\leq N$.
The details of the calculation are shown in Appendix \ref{sec:app_eigenfunction}.
The global power bound (\ref{eq:normalization_constant}) becomes
\begin{align}
 \mathcal{P}_0=\frac{1}{4}\sum_{k=-\infty}^\infty |c_k^q|^2  \sum_{i=1}^{N}  \lambda_i (H_0)_i^2, 
\label{eq:P0_eigenfunctions}
\end{align}
which will be used later.

\subsection{Optimal protocol and maximal power}
\label{sec:optimal_power}

Now, in analogy to section \ref{sec:general_opt_protocol}, we 
maximize the power output of the heat engine.
Defining
\begin{align}
 (u_{\alpha \beta})\equiv
		  \begin{pmatrix}
                   (\Lambda-1)\chi^2 & (\Lambda-1)\chi \\
		  \Lambda \chi \bar\eta &  \Lambda \bar\eta
                  \end{pmatrix},
\end{align}
leads to a compact expression for the objective functional (\ref{eq:objective_functional})
\begin{align}
\mathcal{P}[g_w,\Lambda]
=\sum_{\alpha,\beta} u_{\alpha \beta}  \sum_{k=-\infty}^\infty\sum_{i=1}^{N} \frac{-\ic k \Omega \lambda_i }{\lambda_i -\ic k \Omega} c_k^{\alpha_i} c_{-k}^{\beta_i} ,
\label{eq:OF}
\end{align}
where the $\Lambda$ dependence is through $u_{\alpha \beta}$.
Its maximization
yields the optimal protocol $g_w^*(\Lambda)$ in terms of the coefficients
\begin{align}
 (c_k^{w_i})^*(\Lambda)=\frac{c_k^q (H_0)_i}{2 u_{ww}}\left[u_{wq}+u_{qw}+\frac{\ic \lambda_i}{k\Omega}(u_{wq}-u_{qw}) \right]
\label{eq:general_opt_prot_eigenfunctions}
,\end{align}
for $i \leq n$.
The Lagrange multiplier, implicitly contained in $u_{\alpha\beta}$, is still to be determined.
This result is in analogy to the example of an overdamped Brownian particle in \cite{Brandner2015a}.

Then, with the optimal protocol and $\Lambda=0$, the maximum power is found to be
\begin{align}
 \mathcal{P}(\bar\eta_\text{MP})
=\frac{1}{2} \sum_{k=1}^{\infty}  |c_k^q|^2 \sum_{i=1}^{n}   \lambda_i   (H_0)_i^2 ,
\label{eq:maxPower_schmiedl}
\end{align}
with (\ref{eq:OF}) and the definition (\ref{eq:def_maxPower}).
Similarly we obtain
\begin{align}
\mathcal{J}_q^\text{idle}=\sum_{i=n+1}^{N} L_{qq,i},
\end{align}
with $L_{qq}=\sum_{i=1}^{N} L_{qq,i}$
and 
\begin{align}
 L_{qq,i}
 \equiv 2\sum_{k=1}^\infty |c_k^q|^2 \lambda_i\frac{k^2\Omega^2 }{\lambda_i^2 + k^2 \Omega^2}  (H_0)_i^2 .
\label{eq:Lqqmu}
\end{align}
Note that whereas the sum in the maximal power (\ref{eq:maxPower_schmiedl}) ends at $n$,
the sum in the idle heat flux starts at $n+1$.
This shows that $\mathcal{J}_q^\text{idle}$ is the contribution of the eigenfunctions that cannot be controlled to the heat flux.
The maximal power is large if many eigenfunctions can be controlled.
Finally, these results can be inserted in section \ref{sec:FoM} to 
calculate the figure of merit, the maximum efficiency, etc.
The result for the figure of merit is physically intuitive.
A large figure of merit represents a good performance of the engine,
which is achieved by large maximum power $\mathcal{P}(\bar\eta_\text{MP})$ or small idle heat flux $\mathcal{J}_q^\text{idle}$.
Obviously this is possible if the number of eigenfunctions that cannot be controlled $N-n$ is small.

Furthermore, we are interested whether or not the 
global bound on power (\ref{eq:normalization_constant})
can be saturated.
In the best case, i.e., full control ($n=N$),
we generalize the result from Ref. \cite{Brandner2015a} as
\begin{align}
\lim_{\mathcal{ZT}\to\infty}
 \frac{\mathcal{P}(\bar\eta_\text{MP})}{\mathcal{P}_0}
= \frac{2\sum_{k=1}^\infty |c^q_k|^2}{(c_0^q)^2+2\sum_{k=1}^\infty |c^q_k|^2},
\end{align}
where we have used (\ref{eq:P0_eigenfunctions}) and (\ref{eq:maxPower_schmiedl}).
The global bound can only be saturated if
$c_0^q=0$. Since $0\leq \gamma_q(t)\leq 1$, a vanishing coefficient $c_0^q$ leads to $\gamma_q(t)=0$, which makes the engine futile. 
If the control over the system is limited, the maximum power is smaller, leading to a lower degree of saturation of the global bound.

\section{Case study: Underdamped Brownian heat engine in a magnetic field}
\label{sec:example_schmiedl}

\subsection{Model and solution}
Whereas we have excluded an external magnetic field in section \ref{sec:eigenfunctions}, we now present the arguably most simple system including a magnetic field.
We focus on a engine consisting of an underdamped charged Brownian
particle with mass $m$ in a harmonic trap with equilibrium Hamiltonian
\begin{align}
 H_0({\bf x})&=\frac{m}{2}(v_x^2+v_y^2)+\frac{m}{2} \omega_0^2 (x^2+y^2),
\label{eq:H0_schmiedl}
\end{align}
where $\omega_0$ is the trap frequency, $(x,y)$ is the position and $(v_x,v_y)$ the velocity of the particle.
The particle is confined to two dimensions. Perpendicular to this plane, there is a constant magnetic field, which does not contribute to the energy.
We assume that
the strength of the trap is controlled by $\omega(t)=\omega_0+\gamma_{w_1}(t) \Delta \omega$ with small $\Delta \omega$ and time dependence $\gamma_{w_1}(t)$.
This choice leads to $\Delta H=m (\Delta \omega) \omega_0  l_0^2 $ and
\begin{align}
 g_w({\bf x},t)&=\gamma_{w_1}(t)g_{w_1}({\bf x})={\gamma_{w_1}(t)}(x^2+y^2)/{l_0^2},
 \label{eq:gw_schmiedl}
 \end{align}
i.e., $n=1$ in (\ref{eq:gw_zerlegung}),
where we introduced a reference length $l_0\equiv \sqrt{T/m\omega_0^2}$.
It will be crucial to appreciate that we cannot control the kinetic degrees of freedom,
i.e., change the mass of the particle.
We emphasize that (\ref{eq:gw_schmiedl}) is not an eigenfunction of the adjoint Fokker-Planck operator
and therefore the results of section \ref{sec:eigenfunctions} cannot be used.

In equilibrium, the particle obeys underdamped Brownian motion and its dynamics is described by the 
Fokker-Planck-operator \cite{Risken1989}
\begin{align}
 \L_0=\sum_{j=x,y}-\partial_{j} v_j+ \partial_{v_j}\left(\gamma v_j-\frac{F_j}{m}\right)+\frac{\gamma  T}{m}\partial_{v_j}\partial_{v_j},
\label{eq:L0}
\end{align}
with friction constant
$\gamma$.
The external force 
\begin{align}
\begin{pmatrix}
 F_x /m\\ F_y/m
\end{pmatrix}
=
\begin{pmatrix}
\phantom{-} \omega_c v_y -\omega_0^2 x\\
-\omega_c v_x -\omega_0^2 y
\end{pmatrix},
\end{align}
is the sum of the potential gradient and the Lorentz force.
Here we have introduced the cyclotron frequency 
\begin{align}
 \omega_c \equiv {q\mathrm{B}}/{m},
\end{align}
where $\mathrm{B}$ is the (signed) strength of the magnetic field in $z$-direction and $q$ the charge of the particle.
In Appendix \ref{sec:detailed_balance}
we show that the symmetry (\ref{eq:symm_C}) holds in this case study despite the presence of a magnetic field.

To evaluate the Onsager coefficients (\ref{eq:general_OC}), it is useful to find the eigenfunctions of the adjoint Fokker-Planck operator
\begin{align}
 \L_0^\dagger=\sum_{j=x,y} v_j \partial_{j} - \left(\gamma v_j -\frac{F_j}{m}\right)\partial_{v_j} +\frac{\gamma T}{m}\partial_{v_j}\partial_{v_j}
.\end{align}
In particular, we need to know the action of $L_0^\dagger$ on $H_0$ and $x^2+y^2$,
since (\ref{eq:gq}) and (\ref{eq:gw_schmiedl}) involve these terms.
By straight forward calculation, we
find that the adjoint Fokker-Planck-operator has the form
\begin{align}
 \L_0^\dagger=
\begin{pmatrix}
0&0&-\omega_0^2&0&0\\
0&-2\gamma&1&0&0\\
2&-2\omega_0^2&-\gamma&-\omega_c&0\\ 
0&0&\omega_c&-\gamma&0\\
0&\frac{4\gamma  T}{m}&0&0&0
\end{pmatrix},
\label{eq:L0dagger_schmiedl}
\end{align}
on the invariant subspace $ \mathcal B$ with 
\begin{align}
 {\mathcal{B}}=\text{span}\{x^2+y^2,v_x^2+v_y^2,x v_x+y v_y, x v_y-y v_x,1\},
\label{eq:relevant_subspace}
\end{align}
using this basis.
Functions outside $\mathcal B$ will not be needed in this case study.
The matrix (\ref{eq:L0dagger_schmiedl}) can be diagonalized (see Appendix \ref{sec:app_schmiedl}),
which allows us to evaluate the matrix exponential contained in the Onsager coefficients (\ref{eq:general_OC}).

The Onsager coefficients and the optimal protocol of this case study are shown in Appendix \ref{sec:app_schmiedl}.
We observe that, due to the symmetry of the system, the sign of the magnetic field has no effect on the Onsager coefficients.
Thus, time reversal symmetry is only broken by the protocols and
not by the magnetic field.
Proceeding in analogy to section \ref{sec:limitedcontrol_optdriving}, for $\Lambda=0$ 
we find the maximal power and the idle heat flux
\begin{align}
\mathcal{P}(\bar\eta_\text{MP})&= 4\gamma  T^2  A,\label{eq:PA}\\
\mathcal{J}_q^\text{idle}&=8\gamma  T^2 B,
\end{align}
with abbreviations
\begin{align}
A&\equiv \sum_{k=1}^\infty |c_k^q|^2\frac{1}{4+2b_k(1+{4\gamma^2}/{k^2\Omega^2})}\geq 0,\label{eq:A}\\
B&\equiv \sum_{k=1}^\infty |c_k^q|^2\frac{b_k}{4+2b_k(1+{4\gamma^2}/{k^2\Omega^2})}\geq 0,
\end{align}
and
\begin{align}
 b_k&\equiv \frac{k^2 \Omega^2}{2 \omega_0^2}\left(1+\frac{\omega_c^2}{\gamma^2+k^2 \Omega^2}\right)\geq 0.
\label{eq:bn}
\end{align}
In particular, we obtain the figure of merit
\begin{align}
 \mathcal{ZT}=2A/B,
\label{eq:FoM_schmiedl}
\end{align}
which enables access to the results of subsection \ref{sec:FoM}.

Using the adjoint Fokker-Planck operator (\ref{eq:L0dagger_schmiedl}), and performing the average, the
global power bound (\ref{eq:normalization_constant}) reads
\begin{align}
\mathcal{P}_0
 &=\frac{\gamma  T^2}{2}\sum_{k=-\infty}^\infty |c_k^q|^2.
\label{eq:normalization_constant_schmiedl}
\end{align}

\subsection{Sinusoidal temperature protocol}
\label{sec:sinus}

In the following,
we discuss the behavior of the maximum power output for a specific temperature protocol
to study the influence of the strength of the trap $\omega_0$, the magnetic field $\omega_c$, the inverse cycle time $\Omega$, and the friction constant $\gamma$.
For the protocol of the temperature variation, we choose 
a sinusoidal function
\begin{align}
 \gamma_q(t)=({1+\sin \Omega t})/{2},
\end{align}
resulting in three non-vanishing Fourier coefficients (\ref{eq:cn})
\begin{align}
c^q_k=
 \begin{cases} 1/2 &\mbox{if } k=0, \\
-\ic /4 & \mbox{if } k=+1,\\
+\ic /4 & \mbox{if } k=-1,\\
0 & \mbox{else}.
\end{cases} 
\end{align}
From (\ref{eq:opt_prot}) we see that the optimal protocol $\gamma_{w_1}^*(t)$
is a linear combination of $\sin \Omega t$ and $\cos \Omega t$, which we do not show explicitly.
With the dimensionless parameter
$b_1$
from (\ref{eq:bn}),
the figure of merit (\ref{eq:FoM_schmiedl}) becomes
\begin{align}
 \mathcal{ZT}= \frac{2}{b_1}
=\frac{4\omega_0^2}{\Omega^2}\frac{\gamma^2+\Omega^2}{\gamma^2+\Omega^2+\omega_c^2}.
\label{eq:ZT_sin}
\end{align}
Using this expression, we can qualitatively infer how the system parameters affect the heat engine.
The figure of merit must be large for a good performance,
which is achieved by large strength of the trap $\omega_0$, small magnetic field $\omega_c$
and slow driving, i.e., small cycle frequency $\Omega$.

The power output (\ref{eq:general_power}) becomes
\begin{align}
\frac{\mathcal{P}(\bar\eta)}{\mathcal{P}_0}=\frac{2}{3} \bar\eta \frac{ 1-\bar\eta +{2}/{\mathcal{ZT}} +\sqrt{(1-\bar\eta)^2-4{\bar\eta}/{\mathcal{ZT}}}   }{1+(1+4\gamma^2/\Omega^2)/\mathcal{ZT}}
\label{eq:power_sin}
,\end{align}
where we have used (\ref{eq:PA}) and
$\mathcal{P}_0=3  \gamma  T^2/16 $ from (\ref{eq:normalization_constant_schmiedl}).
Whereas in the previous subsection we normalized power by its maximum,
here we normalize it by the global bound on power (\ref{eq:normalization_constant})
to show the degree of saturation of inequality (\ref{eq:bound_on_power}).
Thus, in (\ref{eq:power_sin}) we have expressed the maximal power for fixed efficiency $\bar\eta$, in terms of cycle frequency in units of the friction constant $\Omega/\gamma$ and $\mathcal{ZT}$.
Despite the fact that $\Omega/\gamma$ and $\mathcal{ZT}$ are not independent, this choice is suitable for the analysis of the power, since
even for fixed $\Omega/\gamma$, the figure of merit (\ref{eq:ZT_sin}) can still obtain all positive values,
which can, for instance, be achieved by changing the strength of the trap $\omega_0$
and the magnetic field $\omega_c$.

\begin{figure}
\includegraphics{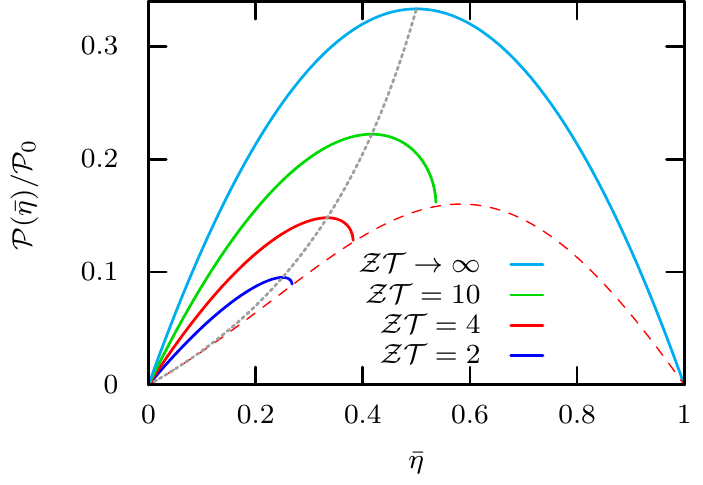}
\caption{Maximum power of the Brownian heat engine (\ref{eq:power_sin}) as a function of the efficiency (solid lines) for sinusoidal driving. From bottom to top the curves are obtained by increasing $\mathcal{ZT}$ and setting $\Omega/\gamma=1$. The dashed line is the power at maximal efficiency (\ref{eq:power_at_max_eff_sinus}) for $\Omega/\gamma=1$, independent of $\mathcal{ZT}$. The dotted line describes the maximum of the curves (\ref{eq:max_power_sinus}), reaching $1/3$ for $\mathcal{ZT}\to\infty$.}
\label{fig:konstOmega}
\end{figure}

The maximum power as a function of efficiency is shown in figure \ref{fig:konstOmega}, where we set $\Omega/\gamma=1$ and vary $\mathcal{ZT}$.
The power grows with a larger figure of merit. For finite $\mathcal{ZT}$, the maximal efficiency is below the Carnot value.
A small value for the figure of merit (\ref{eq:ZT_sin}) means that kinetic effects
play a significant role for the performance of the engine. 
They were introduced via the velocity dependent term in the equilibrium Hamiltonian (\ref{eq:H0_schmiedl})
and influence the Fokker-Planck dynamics (\ref{eq:L0}). However, they are not affected by the modulation of the trap (\ref{eq:gw_schmiedl}),
and
therefore the control on the system is limited.
For instance, the kinetic effects dominate if the strength of the trap $\omega_0$ is small or the magnetic field $\omega_c$ is large.
In the case $\mathcal{ZT}\to\infty$ (see section \ref{sec:tight_coupling}), which corresponds to the overdamped regime (see the example in Ref. \cite{Brandner2015a}),
the kinetic effects vanish and the maximum power is largest.

\begin{figure}
 \includegraphics{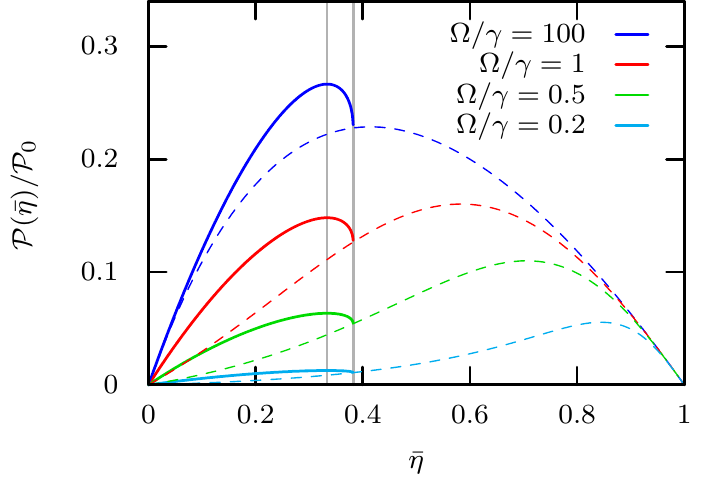}
\caption{Maximum power of the Brownian heat engine (\ref{eq:power_sin}) as a function of the efficiency (solid lines) for sinusoidal driving. From bottom to top the curves are obtained by increasing $\Omega/\gamma$ and setting $\mathcal{ZT}=4$. The dashed lines are the power at maximal efficiency (\ref{eq:power_at_max_eff_sinus}) for different $\Omega/\gamma$. The vertical lines mark maximum efficiency (\ref{eq:max_eff}) and efficiency at maximum power (\ref{eq:EMP}), which both only depend on $\mathcal{ZT}$.}
\label{fig:konstb}
\end{figure}

The maximum power irrespective of $\bar\eta$ reads from (\ref{eq:PA})
\begin{align}
 \frac{\mathcal{P}(\bar\eta_\text{MP})}{\mathcal{P}_0}=\frac{4}{3}\frac{\bar\eta_{\text{MP}}}{4\bar\eta_{\text{MP}}+(1+4\gamma^2/\Omega^2)(1-2\bar\eta_{\text{MP}})},
\label{eq:max_power_sinus}
\end{align}
where we have replaced $\mathcal{ZT}$ via the efficiency at maximum power (\ref{eq:EMP}).
This quantity is shown as a dotted gray line in figure \ref{fig:konstOmega}.
By substituting (\ref{eq:max_eff}) into (\ref{eq:power_sin}), we find the power at maximal efficiency
\begin{align}
 \frac{\mathcal{P}(\bar\eta_\text{max})}{\mathcal{P}_0}=\frac{4}{3}\frac{\bar\eta_\text{max}(1-\bar\eta_\text{max}^2)}{4\bar\eta_\text{max}+(1+4\gamma^2/\Omega^2)(1-\bar\eta_\text{max})^2}
\label{eq:power_at_max_eff_sinus}
,\end{align}
where we have used (\ref{eq:max_eff}) again to replace $\mathcal{ZT}$ by the maximum efficiency.
This result is shown in figure \ref{fig:konstOmega} and \ref{fig:konstb}.
In the latter figure, $\mathcal{ZT}=4$ is fixed
and $\Omega/\gamma$ is varied, which corresponds to changing the inverse cycle time.
Quite naturally, a faster cycle leads to higher power output. Since $\mathcal{ZT}$ is kept constant, maximum efficiency (\ref{eq:max_eff}) and efficiency at maximum power (\ref{eq:EMP}) are fixed.

In our case study, we find at most
the same degree of saturation of the bound on power (\ref{eq:bound_on_power}) as in Ref. \cite{Brandner2015a}
($\mathcal{P}(\bar\eta)/\mathcal{P}_0\leq 1/3$) for a sinusoidal protocol. The fact that the kinetic terms cannot be controlled in the underdamped case,
leads to a degree of saturation of the bound on power even less than $1/3$.

\section{Conclusion}
\label{sec:conclusion}

In summary, we have investigated the performance of periodically driven, stochastic heat engines in the linear response regime under optimal driving.
Our results make predictions for engines, which obey Fokker-Planck dynamics and for which the detailed balance condition leads to symmetric correlation functions.
In particular, if the control of the system is limited, we have shown that the efficiency is bounded, where the bound is described by a single parameter,
the figure of merit. 
It is intuitively clear that less control of an engine may lead to a lower power output,
but our approach treats this problem quantitatively.
The power-efficiency curves take a loop-shaped form, in analogy to imperfect, macroscopic heat engines.
The results are in strong analogy to the linear thermodynamics of thermoelectric steady state devices,
where time reversal symmetry can be broken by an external magnetic field.
In our set-up, time reversal symmetry is additionally broken by the protocols.
Our findings demonstrate the consequences of the framework from Ref. \cite{Brandner2015a} on the thermodynamics of heat to work conversion
under optimal conditions. 

The identification of the figure of merit is only possible if detailed balance can be exploited to show
that the correlation functions are symmetric, which directly follows for systems without magnetic field.
We have, however, identified under which conditions this symmetry holds even in the presence of a magnetic field.
In section \ref{sec:eigenfunctions}, we have shown that the figure of merit is closely related to
the limitations on the control of the system. In a case study, we have investigated
a paradigmatic underdamped example with magnetic field. Due to the symmetry of the system,
the general theory of section \ref{sec:limitedcontrol_optdriving} applies
and we find that the kinetic effects present in the system decrease the performance of the engine.

For a divergent figure of merit, which corresponds to the endoreversible limit of macroscopic engines,
the power output is maximal. 
The optimal protocol shows that full control is needed for the engine to operate at Carnot efficiency,
which comes with vanishing power.
Then the engines fulfills the tight-coupling condition and the Onsager matrix
is symmetric.
Similar results were found in Refs. \cite{Proesmans2015a,Proesmans2015c},
where
the zero dissipation limit of periodically driven systems obeying a master equation was investigated.
However, therein the control of the system is not limited in the sense of (\ref{eq:gw_zerlegung})
and the optimization is not of a functional type.
We expect that the condition (\ref{eq:Carnot_condition}) is equivalent to the ``global modulation of the energy levels'' in Ref. \cite{Proesmans2015a}.

The one-parameter description found here differs from the two parameter description introduced in Ref. \cite{Benenti2011}
for steady state engines with a magnetic field.
In their case the Onsager coefficients are fixed and the optimization refers to the affinities.
In our case we optimize the protocols entering the Onsager coefficients,
which effectively implies a larger variational space.

For future research on the theoretical side,
it will be of interest to investigate
whether the simple relation between
maximum efficiency and efficiency at maximum power (\ref{eq:master_curve})
can be deduced directly from a hidden, underlying concept.
Second, one should investigate whether there exists an example 
that violates the symmetry of the correlation function (\ref{eq:symm_C}).
In such a case a two-parameter description similar to Ref. \cite{Benenti2011} might be necessary.
Third, our framework might be extended to engines beyond Fokker-Planck dynamics, e.g., to quantum systems obeying a Lindblad equation.
Finally, on the experimental side,
it would be interesting to measure a bound on efficiency if there is a constraint on the control of the engine,
which seems to be feasible with regard to recent single particle experiments \cite{blic12,ross15,mart15,mart16}.

\begin{acknowledgements}
We acknowledge stimulating discussions with K. Saito.
\end{acknowledgements}

\appendix

\section{Symmetries of the correlation function implied by detailed balance}
\label{sec:detailed_balance}

We investigate the implications
of detailed balance (\ref{eq:detailed_balance})
on the correlation function (\ref{eq:def_corr}),
where insertion yields
\begin{align}
C[R({\bf x}) ,S({ \bf x}) ,t]
&=\langle R({\bf x}) \e^{\L_0^\dagger({\boldsymbol\varepsilon \bf x}) t} S({\bf x}) \rangle_{\bf x},\nonumber\\
&=\{ C[S({\boldsymbol\varepsilon \bf x}) ,R({\boldsymbol\varepsilon\bf x}),t] \}^{\B\to -\B}.
\label{eq:corr_DB}
\end{align}
In the whole bracket the magnetic field is inverted.
In the last step, we have changed the sign of all odd variables, leaving the area of integration invariant.
The arguments of the correlation functions above are $R({\bf x}),S({\bf x})=\delta H_0({\bf x}),\delta g_{w_i}({\bf x})$.
The energy does not change under inversion $\boldsymbol \varepsilon$
\begin{align}
H({\bf \boldsymbol\varepsilon x}, t)=H({\bf x}, t),
\label{eq:symmetry_Hamiltonian}
\end{align}
for any time $t$.
Then, using (\ref{eq:energy_modulation}),
we find $H_0({\bf \boldsymbol\varepsilon x})=H_0({\bf x})$, $g_{w}({\bf \boldsymbol\varepsilon x},t)=g_{w}({\bf x},t)$,
and also $g_{w_i}({\bf \boldsymbol\varepsilon x})=g_{w_i}({\bf x})$,
since in (\ref{eq:gw_zerlegung}) the functions $\gamma_{w_i}(t)$ are arbitrary.
Using these symmetries, (\ref{eq:corr_DB}) reads
\begin{align}
 C[R({\bf x}) ,S({\bf x}) ,t]
&=\{ C[S({\bf x}) ,R({\bf x}),t] \}^{\B\to -\B}.
\label{eq:corr_DB2}
\end{align}
Obviously, in systems without external magnetic field, we have thus shown that symmetry (\ref{eq:symm_C}) holds, leading to $\mathcal{D}=\mathcal{P}(\bar\eta_\text{MP})$.

If an external magnetic field is present, we can argue as follows.
If the relation
\begin{align}
\Big(\L_0^\dagger({\bf x})-\{\L_0^\dagger({\bf x})\}^{\B\to-\B} \Big) R({\bf x})=0,
\label{eq:symm_cond_with_Bfield}
\end{align}
holds, the required symmetry (\ref{eq:symm_C}) follows from (\ref{eq:corr_DB2}),
since the Hamiltonian $H_0({\bf x})$ and the control functions $\delta g_{w_i}({\bf x})$ do not depend on the magnetic field.
In most systems 
friction and diffusion constants are independent of
the magnetic field. The magnetic field only enters via the Lorentz force.
For example, for one particle in three dimensions, 
\begin{align}
\L_0^\dagger({\bf x})-\{\L_0^\dagger({\bf x})\}^{\B\to-\B}  &=\frac{2q}{m}\sum_{j=x,y,z}(\vec v \times \vec \B)_j \partial_{v_j},\\
&=2\omega_c (v_y \partial_{v_x}-v_x\partial_{v_y}).
\end{align}
In the last step we have assumed that the magnetic field is in $z$-direction only.
If the $\delta g_{w_i}({\bf x})$ are symmetric in the components of the velocity $v_x$ and $v_y$,
then (\ref{eq:symm_cond_with_Bfield}) indeed vanishes.
This reasoning is valid for the case study in section \ref{sec:example_schmiedl}.

In section \ref{sec:eigenfunctions}, we have used
the eigenfunctions of the adjoint Fokker-Planck operator
as control functions $\varphi_i^\dagger({\bf x})= g_{w_i}({\bf x})$.
Above we have shown that the latter are symmetric under $\boldsymbol \varepsilon$,
which is in general not the case for the eigenfunctions.
In the underdamped case, the coefficients $\gamma_{w_i}(t)$ and $(H_0)_i$ have to restore the $\boldsymbol \varepsilon$-symmetry of $H_0({\bf x})$ and $g_w({\bf x})$
in (\ref{eq:gw_zerlegung}) and (\ref{eq:H0_control}).
When performing the optimization of the power output,
we have ignored this demand,
i.e., we have assumed that the Fourier coefficients
$c^{w_i}_k$ are independent of each other.
For our purpose, it is sufficient to show that the optimal protocol (\ref{eq:general_opt_prot_eigenfunctions}) guarantees the symmetry $g _{w}({\bf x})=g_{w}({\boldsymbol \varepsilon \bf x})$.
Due to the summation in (\ref{eq:gw_zerlegung}), we find that the optimal protocol has two contributions in phase space:
\begin{align}
 \sum_{i=1}^{n}(H_0)_i \varphi_i^\dagger({\bf x})&=  H_0^c({\bf x}),\\
 \sum_{i=1}^{n}\lambda_i(H_0)_i \varphi_i^\dagger({\bf x})&=-\L_0^\dagger({\bf x})  H_0^c({\bf x}).
\end{align}
If $n=N$, the first term is $\boldsymbol\varepsilon$-symmetric due to (\ref{eq:symmetry_Hamiltonian}).
For the second term, the adjoint Fokker-Planck operator is divided in a reversible and an irreversible contribution (see Ref. \cite{Brandner2015a}),
where the former is odd and the latter is even under $\boldsymbol\varepsilon$.
The reversible contribution applied on $H_0({\bf x})$ vanishes since it preserves the energy. Then, the second term above is also symmetric under $\boldsymbol\varepsilon$.
If $n<N$, it must be assumed that the same reasoning holds for $H_0^c({\bf x})$.

In summary, we have thus argued that the symmetry (\ref{eq:symm_C})
is fulfilled for a large class of systems, which motivates to focus on the case $\mathcal{D}=\mathcal{P}(\bar\eta_\text{MP})$.

\section{Eigenfunction expansion}
\label{sec:app_eigenfunction}

In general the Fokker-Planck operator
$\L_0({\bf x})$ is non-Hermitian.
We assume that a set of eigenfunctions exists
\begin{align}
 \L_0({\bf x}) \varphi_\mu({\bf x})&=-\lambda_\mu  \varphi_\mu({\bf x}),\\
 \L_0^\dagger({\bf x}) \varphi_\mu^\dagger({\bf x})&=-\lambda_\mu  \varphi_\mu^\dagger({\bf x}),
\end{align}
with
$\lambda_\mu,\varphi_\mu({\bf x}),\varphi_\mu^\dagger({\bf x}) \in \mathbb{C}$, 
which form an
orthonormal set $\int \dd^d {\bf x} \varphi_\mu({\bf x})\varphi_\nu^\dagger({\bf x})=\delta_{\mu \nu}$,
where $\delta_{\mu \nu}$ is the Kronecker delta.
For $\mu \neq 0$, we find $\delta \varphi_\mu^\dagger({\bf x})=\varphi_\mu^\dagger({\bf x})$ due to the orthogonality.
Since the Fokker-Planck operator is real,
for real eigenvalues the corresponding eigenfunctions are also real.
For complex $\lambda_\mu$, its complex conjugate is also an eigenvalue and the corresponding eigenfunctions are conjugate.
It is shown in Ref. \cite{Risken1989} that
$\mathrm{Re} \lambda_\mu >0$ ($\mu \neq 0$),
and therefore each initial distribution approaches equilibrium.

Applying detailed balance (\ref{eq:detailed_balance})
\begin{align}
 \L_0 ({\bf x}) p_{\text{eq}}({\bf x}) \varphi_\mu^\dagger({\bf \boldsymbol\varepsilon x}) &=p_{\text{eq}}({\bf x})\L_0^\dagger ({\bf \boldsymbol\varepsilon x})  \varphi_\mu^\dagger({\bf \boldsymbol\varepsilon x})\\
&=-\lambda_\mu^{\B\to-\B} p_{\text{eq}}({\bf x}) \varphi_\mu^\dagger({\bf \boldsymbol\varepsilon x}),
\end{align}
shows that
\begin{align}
\varphi_\mu({\bf x})=
 p_{\text{eq}}({\bf x}) \varphi_\mu^\dagger({\bf \boldsymbol\varepsilon x}) 
\label{eq:eigenfunction_scaling}
,\end{align}
where in the last step it is crucial to assume the absence of an external magnetic field
for the identification of the eigenfunctions.
Then we obtain the relation
\begin{align}
&\avg{\varphi_\mu^\dagger ({\bf \boldsymbol\varepsilon x})  ( \L_0^\dagger)^n       \varphi_\nu^\dagger({\bf x})}\\
&=(-\lambda_{\nu})^n \int \dd^d {\bf x} \varphi_\mu^\dagger ({\bf \boldsymbol\varepsilon x}) p_{\text{eq}}({\bf x})   \varphi_\nu^\dagger({\bf x})\\
&=(-\lambda_{\nu})^n \delta_{\mu\nu},
\label{eq:orthogonality_N}
\end{align}
for
$n=0,1,\dots$.
This orthogonality (\ref{eq:orthogonality_N}) is now used in the calculation of the Onsager coefficients 
$ L_{\alpha \beta}$ (\ref{eq:general_OC}), where we insert $ g_\alpha ({\bf  x},t)$ and $ g_\beta ({\bf \boldsymbol\varepsilon x},t)$.
This procedure yields
\begin{align}
 L_{\alpha \beta}^\text{ad}=\sum_{k=-\infty}^\infty&\sum_{i=1}^{N} -\ic k \Omega c_k^{\alpha_i} c_{-k}^{\beta_i} ,\\
 L_{\alpha \beta}^\text{dyn}=\sum_{k=-\infty}^\infty&\sum_{i=1}^{N} k^2 \Omega^2 c_k^{\alpha_i} c_{-k}^{\beta_i}  \times \\
&\left(\int_0^\infty \dd \tau e^{-\lambda_\mu\tau} e^{\ic k\Omega \tau}  \right)\nonumber\\
=\sum_{k=-\infty}^\infty&\sum_{i=1}^{N} \frac{k^2 \Omega^2}{\lambda_\mu -\ic k \Omega} c_k^{\alpha_i} c_{-k}^{\beta_i} ,
\end{align}
which is combined to the final result (\ref{eq:OC_eigenfunctions}).

\section{Onsager coefficients and optimal protocol for the case study in section \ref{sec:example_schmiedl}}
\label{sec:app_schmiedl}

To calculate phase space averages appearing in the Onsager coefficients (\ref{eq:general_OC}), we need the
equilibrium distribution
\begin{align}
 p^{\text{eq}}({\bf x})=\sqrt{\frac{\Det \sigma^{-1}}{2\pi }}\exp\Big(-\frac{1}{2}\sum_{j,k}x_j (\sigma^{-1})_{jk}x_k \Big),
\end{align}
with $j,k\in \{1,2,3,4\}$,
${\bf x}=(x_1,x_2,x_3,x_4)=(x,y,v_x,v_y)$, $\Det \sigma^{-1}=({m\omega_0}/{ T})^4$,
and the non-vanishing matrix elements
\begin{align}
(\sigma^{-1})_{11}&=(\sigma^{-1})_{22}=m\omega_0^2 /T,\\
(\sigma^{-1})_{33}&=(\sigma^{-1})_{44}=m/T.
\end{align}
The matrix $\sigma$ can easily be found by inversion, which we use to evaluate
second moments $\avg{x_j x_k}=\sigma_{jk}$
and fourth moments  $\avg{x_i x_j x_k x_l}=\sigma_{ij}\sigma_{kl}+\sigma_{ik}\sigma_{jl}+\sigma_{il}\sigma_{jk}$ \cite{Risken1989}.
Higher moments are not needed in this case study.

The adjoint Fokker-Planck operator (\ref{eq:L0dagger_schmiedl}) can be diagonalized,
which leads to the diagonal matrix
\begin{align}
 D
 &= \mathcal{E}^{-1} \L_0^\dagger  \mathcal{E}\nonumber\\
&=\gamma\, \text{Diag}(0,-1-i s_1,-1+i s_1,-1- s_2,-1+ s_2),
\label{eq:diagonal_FP}
\end{align}
where we have used the abbreviations
\begin{align}
 s_0& \equiv \sqrt{1+2(\tilde\omega_c^2-\tilde\omega^2)+(\tilde\omega^2+\tilde\omega_c^2)^2},\\
s_1& \equiv \sqrt{(-1+s_0+\tilde\omega^2+\tilde\omega_c^2)/2},\\
s_2& \equiv \sqrt{(1+s_0-\tilde\omega^2-\tilde\omega_c^2)/2},
\end{align}
and $\omega \equiv 2\omega_0$.
Tilde means division by $\gamma$, resulting in dimensionless frequencies.
For diagonalization
we have used
\begin{widetext}
\begin{align}
 \mathcal{E}=
\left(
\begin{array}{ccccc}
 0 &  (-1+i s_1)  \omega_0^2   &  (-1-i s_1)  \omega_0^2
  & (-1+s_2)  \omega_0^2 & (-1-s_2)   \omega_0^2 \\
 0 & (-1-i s_1)  & (-1+i s_1)  &
   (-1-s_2)  & (-1+s_2)  \\
 0 & -s_3^2\gamma & -s_3^2\gamma & -s_4^2\gamma & -s_4^2\gamma \\
 0 & -i s_3^2 \omega_c/s_1 & i s_3^2 \omega_c/s_1 & s_4^2 \omega_c/s_2 & -s_4^2
   \omega_c/s_2 \\
 1 & 4/m\beta & 4/m\beta & 4/m\beta &
   4/m\beta \\
\end{array}
\right)
,\end{align}
\end{widetext}
with
$s_3^2 \equiv (1+s_0+\tilde\omega^2+\tilde\omega_c^2)/2$ and
$s_4^2 \equiv (1-s_0+\tilde\omega^2+\tilde\omega_c^2)/2$.
Then we rewrite $\exp(\L_0^\dagger \tau)= \mathcal{E} \exp(D \tau) \mathcal{E}^{-1}$.
The columns of $\mathcal{E}$ are eigenvectors of (\ref{eq:L0dagger_schmiedl})
and scalar multiples of the $\varphi_\mu^\dagger(\bf x)$ (see Appendix \ref{sec:app_eigenfunction}).
These eigenfunctions are uniquely determined by the orthogonality relation and (\ref{eq:eigenfunction_scaling}).

With the Fourier series expansion
of $\gamma_{q,w_1}(t)$
and by performing the phase space average,
from (\ref{eq:general_OC}) we find
\begin{align}
 L^{\text{ad}}_{qw}&=- L^{\text{ad}}_{wq}\\
&=\frac{2 T^2}{m\omega_0^2 l_0^2{\mathcal T}} \int_0^{\mathcal{T}} \dot{\gamma}_q(t) \gamma_{w_1}(t) \dd t\\
&=\frac{2 T^2}{m\omega_0^2 l_0^2} \sum_{k=-\infty}^\infty \ic k \Omega c_k^q c_{-k}^{w_1},
\end{align}
and $L^{\text{ad}}_{qq}= L^{\text{ad}}_{ww}=0$, since the $\gamma_{q,w_1}(t)$ are $\mathcal{T}$-periodic.
With the diagonal form of $\L_0^\dagger$, the $\tau$-integration in (\ref{eq:general_OC}) can be performed leading to the second contribution to the Onsager coefficients
\begin{widetext}
\begin{align}
L_{qq}^\text{dyn}&=2\gamma  T^2\sum_{k=1}^\infty |c^q_k|^2  k^2\tilde\Omega ^2\frac{f_k}{r_k},\\
L_{qw}^\text{dyn}&=\frac{2\gamma   T^2  }{  m \omega_0^2 l_0^2}\sum_{k=1}^\infty c^q_{-k} c^{w_1}_k k^2\tilde\Omega ^2 \frac{-f_k-(1+\ic k\tilde\Omega)^2-\tilde\omega_c^2}{r_k},\\
 L_{ww}^\text{dyn}&=\frac{4\gamma  T^2}{
    m^2 \omega_0^4 l_0^4}\sum_{k=1}^\infty
|c^{w_1}_k|^2 k^2\tilde\Omega ^2\frac{f_k+\tilde\omega_c^2+(1+\ic k\tilde\Omega)^2-\tilde\omega^2(1+\ic k\tilde\Omega)/2}
{ r_k}.
\end{align}
\end{widetext}
Here, we have used the abbreviations
\begin{align}
f_k&\equiv  (1+\ic k \tilde\Omega ) \left[\tilde\omega_c^2+\tilde\omega^2+(1+\ic k \tilde\Omega
   )^2\right],\\
r_k&\equiv (1+\ic k \tilde\Omega)^4 +(1+\ic k \tilde\Omega)^2 \left(\tilde\omega_c^2+\tilde\omega^2-1\right)-\tilde\omega_c^2.
\end{align}
The still missing Onsager coefficient $L_{wq}^\text{dyn}$ can be obtained from $L_{qw}^\text{dyn}$ by interchanging the $w_1$- and $q$-protocol. 
This fact originates from the "additional symmetry relation" in Ref. \cite{Brandner2015a}, since $(\ref{eq:gw_schmiedl})$ factorizes.

After a rather lengthy calculation, the conditions $\partial_{c^{w_1}_k}\mathcal{P}[g_w,\Lambda]=0$ with (\ref{eq:objective_functional}) yield the components of the optimal protocol
\begin{widetext}
 \begin{align}
\frac{2\chi}{m\omega_0^2l_0^2} (c_k^{w_1})^*(\Lambda)&=-c_k^q\frac{a_-[2k^2\tilde\Omega^2 \tilde\omega_0^2-k^4\tilde\Omega^4+b_k \tilde \omega_0^2(2+k^2\tilde\Omega^2)]   +\ic a_+ k\tilde\Omega [(2+b_k)\tilde\omega_0^2 -2k^2\tilde\Omega^2]}{\tilde \omega_0^2[2k^2\tilde \Omega^2+ b_k (4+k^2\tilde\Omega^2)]},
\label{eq:opt_prot} 
\end{align}
\end{widetext}
with
$a_\pm\equiv (1-\Lambda\pm \bar\eta\Lambda) / (\Lambda-1)$.
An expansion of (\ref{eq:opt_prot}) in small $\Omega$ is similar
to the optimal protocol found in Ref. \cite{Brandner2015a}.

%

\end{document}